\documentstyle[aps,graphicx,times,floats]{revtex}
\begin{document}
\def\lsco15{La$_{0.85}$Sr$_{0.15}$CuO$_{4}$\ }

\def\ybco6{YBa$_{2}$Cu$_{3}$O$_{6.6}$\ }

\def\ybco7{YBa$_{2}$Cu$_{3}$O$_{7}$\ }
\draft

\wideabs{

  \title{Frequency Evolution of Neutron Peaks Below $T_c$:
Commensurate and Incommensurate Structure in LaSrCuO
and \mbox{YBaCuO} }

  \author{Ying-Jer Kao}

  \address{James Franck Institute, University of Chicago, Chicago, Illinois
    60637, USA}
 
   \author{Qimiao Si}

   \address{ Department of Physics, Rice University, Houston, TX 77251,USA}

   \author{K. Levin}

   \address{James Franck Institute, University of Chicago, Chicago, Illinois
    60637, USA}
  \date{\today} 

\maketitle
%----------------------------------------------------------------
\begin{abstract}
  We study the evolution of the neutron cross-section with variable
  frequency $\omega$ and fixed $T$ below $T_c$ in two different cuprate
  families. Our calculations, which predominantly probe the role of
  $d$-wave pairing, lead to 
  generic features,
  independent of Fermi surface shapes.
  Among our findings, reasonably consistent with experiment, are (i) for
  $\omega $ near the gap energy $\Delta$, both optimal
  {LaSrCuO} and
  slightly underdoped YBCO exhibit (comparably) incommensurate peaks
  (ii) peak sharpening below $T_c$ is seen in {LaSrCuO},
  (iii) quite generically, a frequency evolution from incommensurate to
  commensurate and then back to incommensurate structure is
  found with increasing $\omega$. Due to their narrow $\omega$ regime
  of stability, commensurate
  peaks in {LaSrCuO} should be extremely difficult to observe.

\end{abstract}

}

%------------------------------------------------------
The field of neutron scattering in high temperature superconductors
has seen 
a variety of recent experimental discoveries 
associated with incommensurate and commensurate structure at $T < T_c$
\cite{Mook,Japanese,newKeimer,Aeppli,newMook}.
These neutron data remain of central importance in the field of
high $T_c$ superconductivity: the combined momentum and frequency
range covered by this technique is wider than that
in virtually all other
spectroscopies. It is mainly through this extended range that
these neutron results have direct bearings on such important issues
as dynamical stripes\cite{stripes} and where the condensation
energy comes from\cite{SO5,Zhang2}.

The goal of the present paper is to systematically
address these observations over the \textit{entire} momentum and frequency
range that the experiments have covered, in the two
different cuprate families (\mbox{LaSrCuO} and \mbox{YBaCuO}).  In the process,
we
show that all of the above commensurate and
incommensurate features are compatible with
$d$-wave pairing superposed onto the normal state 
Fermi surfaces. 
In contrast to $T > T_c$, the
details of the Fermi surface shape are of relatively less importance,
and serve primarily to select out the $\omega$ regime where
various commensurate or incommensurate features can be observed.
The importance of the present work derives from the panoply
of different experimental observations which are semi-quantitatively
addressed here.  These calculations have no adjustable
parameters(besides those which were used originally\cite{Si,Zha} to fit
some aspects of normal state data), so that their success
or failure, upon comparison with experiment, should help select
out viable theories of the cuprates.

Here, following previous work above\cite{Si} and
below\cite{Zha,Liu} $T_c$, we apply
an RPA, three band scheme which we have developed to treat the
effects of very strong
Coulomb correlations\cite{earlySi}. 
It should be stressed that our RPA approach is not a weak coupling
RPA. The Lindhard function $\chi_o$ used here is
appropriate whenever
the spin excitations
are associated with spin $1/2$ and with an underlying Fermi surface--
be it in a Fermi liquid\cite{Littlewood,Tremblay,Bulut,Mazin,Bulut2},
or in a spin-charge separated metal\cite{Fukuyama,Lee2,Yin}.
Our starting point is the dynamical susceptibility\cite{earlySi}
  $\chi ( q , \omega ) = \chi ^o ( q , \omega ) / [ 1 + J ( q ) \chi ^o ( q ,
  \omega )]$
where, at low $T$, the dominant contribution to the
imaginary part of $\chi^o$ is given by
\begin{equation}
Im \chi^o(q, \omega)=\sum_{k} u(k, q) \delta(\omega-E_2(k,q))
\end{equation} 
Here $u(k,q)=(1-(\xi_k \xi_{k+q}+\Delta_k \Delta_{k+q})/ 
E_k E_{k+q})$, and $\xi_k$ represents the ``bare" particle
energy relative to the chemical potential while $ E_k$ is
that of the 
superconducting quasi-particles. The important function
$E_2(k,q)= E_k+E_{k+q}$ will play a key role in 
our analysis.

The RPA
neutron cross section reflects a competition
between effects associated with the 
Fermi surface shapes and pairing symmetry [via $\chi_o$]
and those from the residual exchange interaction\cite{earlySi}
$J(q) = J_o [\cos q_x + \cos q_y]$, which derives from
Cu-Cu interactions
via the mediating oxygen band.
While the \mbox{YBaCuO} system is a two layer material, 
our past experience\cite{Si,Liu}
has shown that most of the peak structures associated with the neutron
cross section are captured by an effective one layer band calculation,
which we will investigate here.
For definiteness, we have fixed the temperature at $4$ K 
and assume the electronic excitation
gap to be described by an ideal $d$-wave, $\Delta ( q ) = \Delta (\cos
q_x +\cos q_y)$, where at $ T = 4 $ K, $\Delta$ is taken to be 
$17 $ meV in \mbox{YBaCuO$_{6.6}$} and $8 $ meV in optimally
doped \mbox{LaSrCuO}. 
The breakdown of the Fermi liquid state is addressed only insofar as there
may be precursor pairing or pseudogap effects, which lead to an 
excitation gap in the Lindhard function $\chi_o$ above
$T_c$. Our calculations were based on a numerical procedure in which
the Brillouin zone is sub-divided into tetrahedral
microzones\cite{tetra}. Our peak heights are represented
in arbitrary units, which are best quantified by noting that
in the normal state, the peak values are around 2
for \mbox{LaSrCuO} and for underdoped \mbox{YBaCuO}, whereas for the latter
compound at optimal doping, the value is around 1. Presumably
all peak structure with intensity less than this is not currently observable
(since the normal state peak of \mbox{YBaCuO$_7$} is essentially absent
experimentally\cite{Keimer}).

\begin{figure}
\centerline{
\includegraphics[width=3.5in]{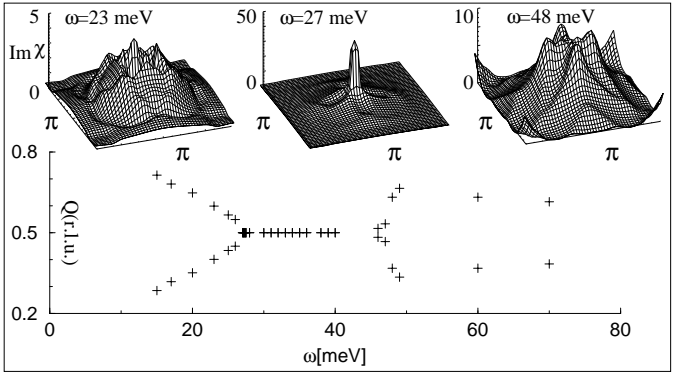}
}
\vskip 4mm
\caption{Frequency evolution of neutron cross section ($Im \chi$)
  for
  \mbox{YBaCuO$_{6.6}$}. Units are arbitrary
  with corresponding normal state values given as a basis in text.
Lower panel indicates the position of the dominant peaks in
reciprical-lattice units (r.l.u.).   }
\end{figure}

In Figure 1 we show the evolution of the neutron peaks in
underdoped \mbox{YBaCuO} as a function of $\omega$. 
These incommensurate peaks are first seen at $ \omega \approx \Delta $;
as frequency increases, the incommensurability is found to
continuously decrease. This decrease is most apparent in
the immediate vicinity of the onset of the resonance,
[or ($\pi , \pi $) peak] which
can be read off from the lower part of Figure 1, to be at around
27 meV, somewhat less than $ 2 \Delta$.  
Just above resonance, the $(\pi , \pi)$ peak becomes flat-topped
possibly weakly incommensurate.
It then broadens and remains structureless between 30-40 meV. Finally,
above 45 meV, clear incommensurate structure re-appears. 
We find that our
low $\omega $ incommensurate peak heights are in the ratio 
of about 1:10 when compared with the resonance feature.  However,
when the integrated spectral weight is considered, the ratio is
about 1:2. Experimentally\cite{Mook}, the ratio of spectral weights is
found to be 1: {3.8} .

The lower
panel shows the detailed frequency evolution of the dominant
peak position
quantified as $(\pi, (1\pm \delta) \pi )=(0.5,0.5 \pm \delta/2)=(0.5,Q)$ 
in reciprocal-lattice units (r.l.u.).
\textit{The peaks evolve 
much as is seen experimentally}\cite{Japanese,newKeimer}. 
The primary difference between our observations
and these particular experiments\cite{Japanese} is that 
over a
range of frequencies, the incommensurate features 
coexist (although, not explicitly indicated in the lower panel)
with the more dominant
resonant peak. By contrast, experimentally, an energy scale
$E_c$ is associated with the frequency at which the various
peaks merge.
It should be stressed, however,
that here we have not incorporated resolution limiting effects
which may affect this detailed comparison between theory and
experiment.
A very early prediction for 
this lowest energy scale
incommensurability was presented by our group
in Ref.\onlinecite{Zha}, where it was
shown to be a consequence of
$d$-wave pairing and
relatively independent of the fermiology. Subsequent
insights, using a related but differently motivated
formalism,  were provided in Ref\onlinecite{Lee2}, which
explicitly showed the influence of $d$-wave pairing on
$ Im \chi_o$.
\begin{figure}
  \centerline{ \includegraphics[width=3.5in]{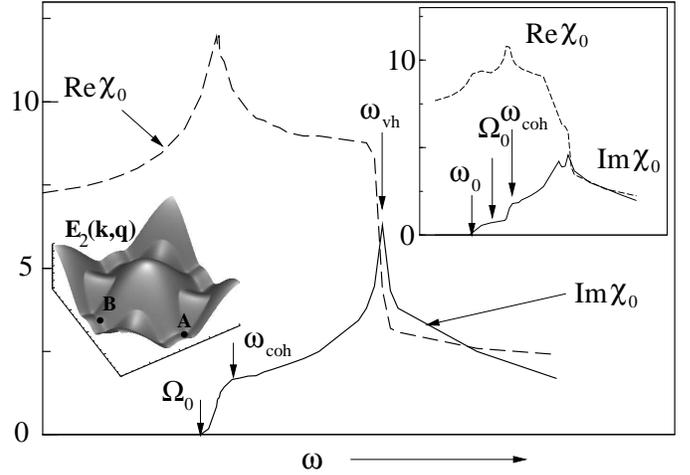} } \vskip 2mm
\caption{Frequency dependence of $Im \chi_o$ and $Re \chi_o$ for
$Q=(\pi,\pi)$.  Upper inset is comparable plot for
$Q=((1+\delta)\pi, \pi)$.  The key characteristic energy
scales (see text) are indicated. Lower (left) inset shows $E_2(k,q=Q_o )$ ; here
A corresponds to the minimum of $E_2(k,q)$, ($\Omega_o$) and B to 
the saddle point of $E_2(k,q)$, ($\omega_{vh}$).
   }
\end{figure}

To understand the origin of the various commensurate and
incommensurate structures, in the main body of Figure 2 we plot the
prototypical behavior of $Re \chi_o$ and $ Im \chi_o$ for
$\bf{q}$ at the \textit{commensurate} point $  Q_o = ( \pi , \pi $); 
in the inset is shown the analogous plot for the
\textit{incommensurate} peaks.  These plots,
which were chosen to correspond to \mbox{YBaCuO}, contain,
in effect, a summary of the key energy scales\cite{Lavagna}
which appear in $ Im \chi_o$. Here we emphasize how
they are reflected in $Re \chi_o$. These
$Re \chi_o$ effects are essential because, through the
RPA denominator they serve to greatly enhance a given characteristic
feature in $Im \chi_o$. 
The four 
important energy scales which determine the behavior of $ Im \chi$
(via simultaneous effects on $ Im \chi_o$ and
$ Re \chi_o$ ) are given as 
(i) $\omega_o({\bf q})= min~ E_2( k, q) $,
(ii) $\omega_{vh}$, the saddle point of $E_2( k, Q_o)$ (see point
B in the inset), (iii) $\omega_{coh}$  the
frequency where the rate of change of $u(k, q)$ drops abruptly.    
and, (iv)
the onset for commensurate peaks $\Omega_o=\omega_o(Q_o)$ (see point A
in the inset).

The related implications for $ Re \chi_o$ are illustrated in
Figure 2. The onset frequency $\Omega_o$
is accompanied
by a substantial growth in $ Re \chi_o$.  However, once the
frequency reaches the two-particle Van Hove energy $\omega_{vh}$,
determined by the saddle point shown as B in
the inset,  
$~ Re \chi_o$ 
shuts down, as does the resonance. 
Figure 2, thus, shows that because
of these $ Re \chi_o$ amplifications along with
the $q$-structure of $J(q)$, the commensurate peak
will tend to dominate incommensurate structure in the range $\Omega_o <
\omega < \omega_{vh}$, because of the small size of the RPA
susceptibility which, in turn, yields a pole-like behavior in $Im \chi$. 
Incommensurate peaks only appear at the lower frequencies because
their threshold $\omega_o$ is less than that of the commensurate
structure. 
The appearance of incommensurability is associated with the
$d$-wave state and is best seen pictorially in the
upper right inset of Figure 3, shown for the
\mbox{LaSrCuO} family which contrasts nesting processes in the
normal state (left) and $d$-wave superconducting state (right). Here the spectral
weight of a given process is indicated by the intensity of the various
lines. The magnitude of the
incommensurability of the peak is determined by (i) energy
conservation through the delta function in Eq.~(1) along with (ii) coherence
factor [$u(k,q)$] effects, which select out the most
favorable regions from otherwise equivalent nesting vectors.

To summarize, we can consolidate our observations, along with a collection
of some of the mechanisms\cite{Mazin,Liu,Keimer2} which have been proposed
for the resonance in YBaCuO into a single inequality: low $\omega$
incommensurate peaks appear for $\omega_o < \omega < \Omega_o $, and
commensurate peaks appear for $\Omega_o < \omega < \omega_{vh}$. At still higher
frequencies the peaks are again incommensurate. $\Omega_o$ is always
somewhat less than $2 \Delta$ because of the nodal structure of the
$d$-wave gap. 

Indeed, the results of Figure 2 are rather generic and can be used to
address the \mbox{LaSrCuO} family as well,
where unusual incommensurate structures in the cross
section have been experimentally
reported\cite{Aeppli} below $T_c$ for a range of low $\omega$,
above $\Delta$. These, and related ``spin gap" features have not
yet been 
addressed theoretically. Here, 
we demonstrate that, in contrast to Ref. \onlinecite{Morr},
these features-- which are closely connected to the (low energy
scale) incommensurate
peaks discussed above for \mbox{YBaCuO}-- 
are unrelated to
fermiology effects and to presumed ``incommensurate
spin structure".
They depend exclusively on the $d$-wave 
pairing symmetry. Figure~3 shows a comparison of the 
cross section in the normal
( left) and superconducting (right) phases of optimally doped
\mbox{LaSrCuO} at low $\omega$. 
The peak sharpening below $T_c$ was first observed
experimentally\cite{Aeppli}. Its origin 
can be traced to the differences (above and below $T_c$)
in
the initial and final scattering processes along constant energy
contours, which contribute to $Im
\chi^o$ at a given point. These processes are shown in the upper panels
of Figure~3
for a particular point $\bf{q}$ indicated on the normal state cross
section. 
\begin{figure}
\centerline{
\includegraphics[width=3in]{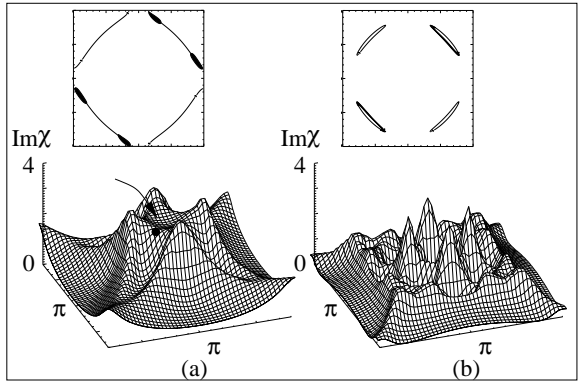}
}
\vskip 4mm
\caption{Origin of peak sharpening in \mbox{LaSrCuO} below $T_c$. Superconducting
  and normal state processes contributing to cross section at indicated
  point are shown in right and left hand figures of upper panel,
  with line
  intensity corresponding to associated weight. Lower panels show
  the corresponding cross sections. 
  }
\end{figure}
It can be seen that in the superconducting phase,
as a consequence of the opening of the gap, 
there is relatively little weight at the $\bf{q}$ point in question,
thereby leading to
a greatly reduced scattering in the
region \textit{between} the incommensurate peaks below $T_c$.

In order to gain more experimental insight into the behavior
of the neutron peaks of \mbox{LaSrCuO} in the superconducting phase,
the authors of Ref. \onlinecite{Aeppli} measured $ Re \chi$
at low frequencies and at the incommensurate points. 
They also
addressed
the details in the frequency
onset of the cross section, measured via $ Im \chi$,
for a range of different wave-vectors. To determine
$ Re \chi$, the cross section was fitted to a simple 
Lorentzian-like form\cite{Aeppli} peaked at $\Delta_s$
Here, we use the directly calculated $Re \chi $ to extract $\Delta_s$.
The resulting experimental curves look qualitatively similar
to the theoretical
plots shown in the right hand inset of Figure 4. A reduction
in $Re \chi( \omega = 0 )$ 
between the normal and superconducting phase
can be associated
(via general Kramers Kronig relations)
with a low $\omega$ suppression in the spectral weight of $ Im \chi$.   
In the present theoretical scheme, this suppression derives from
the opening of a $d$-wave superconducting gap. It is important,
however, to stress that the $d$-wave symmetry in $ Im \chi$
is not immediately evident from either the plots in Figure 4
or the data. Indeed, the main portion of the figure shows
$ Im \chi $ \textit{vs} $\omega$ for three
different values of the incommensurate
wave-vector $ {\bf q} =(1.12h\pi,0.88h\pi)=(0.56h, 0.44h)$ (r.l.u.), 
and indicates that the onset frequency in
$ Im \chi $ is relatively constant in ${\bf q}$.
This onset frequency or `` spin gap" 
$\Delta_s$, is plotted more
completely for a range of wave-vectors in the upper left inset.
Here the range is taken to coincide with its experimental
counterpart and the results shown compare favorably with
experimental data from Ref. \onlinecite{Aeppli}. While
the
calculated spin gap magnitude is larger by a factor of 2 than
experiment, it
should be stressed that the downturn at large $h$ is seen
experimentally, and is predicted to be systematic.

\begin{figure}
\centerline{
\includegraphics[width=3in]{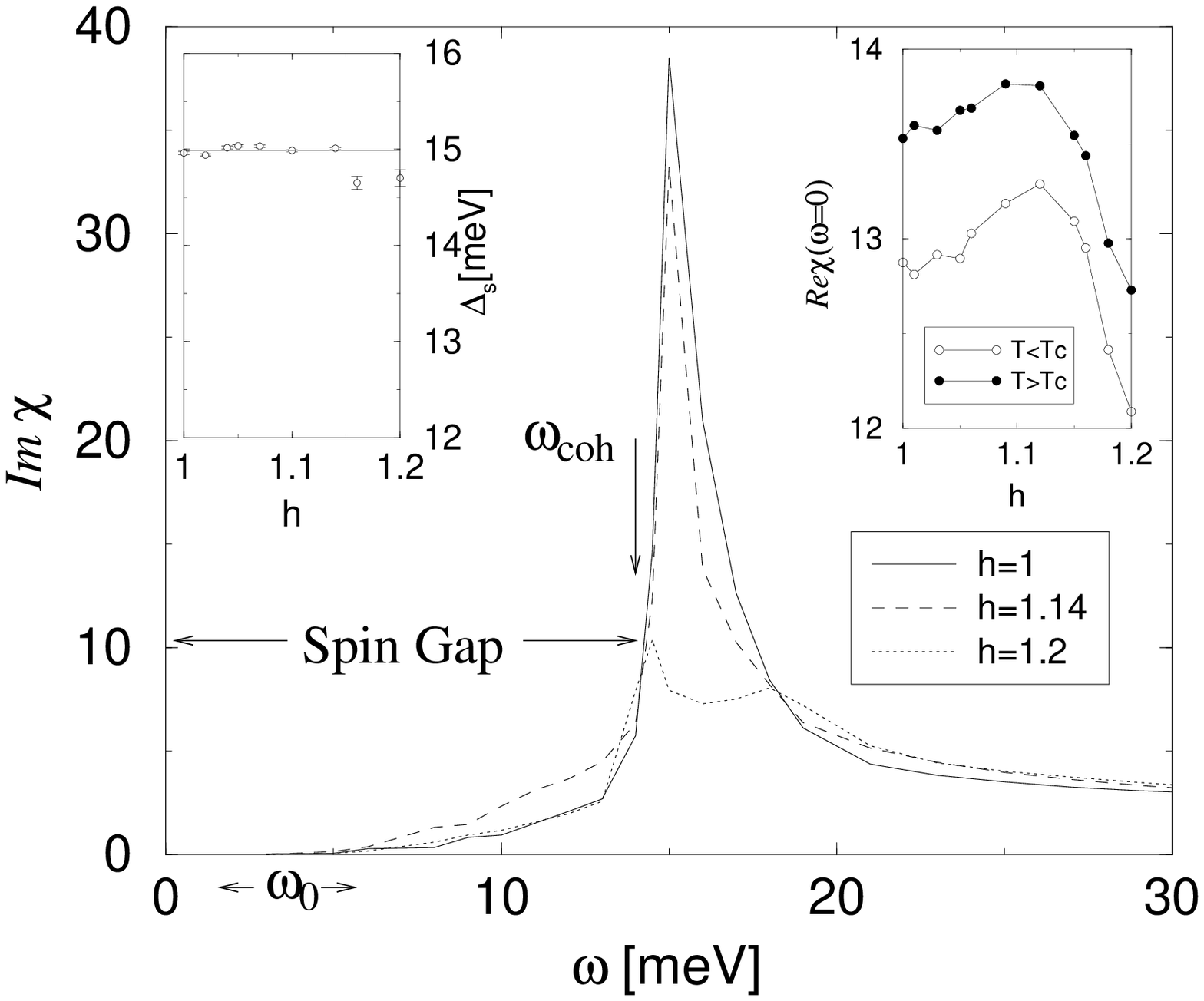}
}
\caption{$Im \chi$ for \mbox{LaSrCuO} in the superconducting
state along ${\bf q}=(0.56h,0.44h)(r.l.u.)$ at $h=1,1.14,1.2$. Lefthand inset shows the ``spin
gap" obtained by fitting $Im \chi$ to a Lorentzian.
Righthand inset
compares $Re \chi$ at $\omega=0$ in the normal and superconducting states. }
    
\end{figure}

Why then is there no $d$-wave signature in the spin gap frequency?
This follows because the $\bf{q}$ structure
associated with the $d$-wave gap appears inside the 
integral over ${\bf k}$ in
the Lindhard function $\chi_o$.
Moreover, the $\bf{q}$ points chosen in the main
body of the figure and in the left hand inset, while consistent
with those measured in Ref. \onlinecite{Aeppli}, do not reflect
most directly the
superconducting gap.  In our calculations
the spin gap $\Delta_s$ is found to be associated with
the coherence
(coh) factor energy
scale, $\omega_{coh}$, which is relatively wave-vector
independent in the region studied. 
However, it is the slightly lower, threshold 
frequency at which $ Im \chi$ first
becomes non-zero ( $\omega_o(q)$), which most closely reflects
$\Delta_q$. (Indeed, $\omega_o$ varies by a factor of 2
for the $q$ values indicated). By contrast
with $ \omega_{coh}$, at $\omega_o$, the cross section has
relatively
little detectable weight.

What, then, are the similarities between the \mbox{LaSrCuO} and \mbox{YBaCuO}
families? There are claims in the literature\cite{Mook,Aeppli}
that the incommensurabilities for near-optimal
\mbox{LaSrCuO} and for \mbox{YBaCu$O_{6.6}$} are the same, as are the spectral
weights once $\omega$ is scaled by the appropriate $T_c$.
To address these more quantitative issues we first note that
the $\omega = 23 $ meV incommensurability shown in underdoped in \mbox{YBaCuO} 
in Figure 1 is comparable, $ \approx 0.1 $ (r.l.u.),
to that plotted in Figure 3 for optimal
\mbox{LaSrCuO}. 

Moreover,
as an estimate of the
q-integrated spectral weight, we
find that the Brillouin zone averages, called
$\langle Im \chi({\bf q},\omega)\rangle_{BZ}$ are equal for 
\mbox{YBaCuO$_{6.6}$} and optimal
\mbox{LaSrCuO} at 22 meV and 13 meV, respectively. If we rescale $\omega$ by $T_c$ in
each case ($T_c \approx 60$K for \mbox{YBaCuO$_{6.6}$}\cite{Dai}, $35$ K 
for La$_{0.85}$Sr$_{0.15}$CuO$_4$\cite{Yamada}), it follows
that at about the same
re-scaled frequencies, these intensities
are nearly equal. It should be stressed that we regard the underlying
physical origin of the  incommensurate peaks \textit{in 
the superconducting state} of the
two cuprates to be the same (as discussed in the context of Figure 2), but that the
characteristic energy and wave-vector scales need not be precisely
equal. Indeed, the incommensurability is now found to be $\omega$
dependent\cite{Japanese}, which makes the numerical
comparisons somewhat
less meaningful.
Finally, it should be noted that the
main body of Figure 4 is rather generic (as seen in Figure 2). These plots
of $ Im \chi $ together with their normal state counterparts, are
in reasonable agreement with the results of Ref. \onlinecite{Aeppli}. 

What then is the most significant difference between the \mbox{LaSrCuO}
and \mbox{YBaCuO} families? We find that this difference derives from
the fine
details of the bandstructure which are then reflected in
the range of stability of commensurate peaks.  
The former compound is presumed
to have a significantly smaller next nearest neighbor hopping integral
($t'$) so that the two particle Van Hove feature\cite{Mazin,Liu} 
$\omega_{vh}$ appears much
closer to $\Omega_o$. Indeed, for \mbox{LaSrCuO} essentially all
energy scales (for commensurate structure) are compressed down to
$\Omega_o$.  
In this way
there is a much smaller frequency window ($\approx$ 2meV)
for commensurate peaks
in the \mbox{LaSrCuO} family than in its \mbox{YBaCuO} counterpart. 
Thus far, searches within the broader 5 meV interval
have failed to see them. 
While this narrow regime of stability will make them difficult
to see, commensurate peaks appear to be a fairly
general feature of the RPA- like $d$-wave approach to the
neutron cross section. 

In summary, it should be stressed that the frequency evolution
from incommensurate to commensurate and then to incommensurate
peaks in the neutron cross section (as shown by the lower panel
in Figure 1) is a generic feature of the
present approach, and because it appears to be observed experimentally
in at least two cuprate families\cite{Mook,newMook}, it will be
important to establish whether an alternative scheme, such
as the ``stripe" picture
can lead to similar behavior.

We would like to thank G. Aeppli for informative and illuminating discussions.
This work was supported by the NSF under awards No.~DMR-91-20000
(through STCS), and No.~DMR-9808595 (through MRSEC). Q.~S. is  supported 
in part by NSF Grant No.~DMR-9712626, Research Corporation, and A. P. Sloan Foundation.

%\bibliographystyle{prsty}
%\bibliography{./neutron}

\begin{thebibliography}{10}

\bibitem{Mook}
{H. A. Mook \textit{et~al.}}, Nature {\bf 395},  395  (1998).

\bibitem{Japanese}
M. Arai {\it et~al.}, Phys. Rev. Lett. {\bf 83},  608  (1999).

\bibitem{newKeimer}
{H.~F. Hong \textit{et~al.}}, cond-mat/9910041 (unpublished).

\bibitem{Aeppli}
B. Lake {\it et~al.}, Nature {\bf 400},  43  (1999).

\bibitem{newMook}
{H.~A. Mook \textit{et~al.}}, cond-mat/9811100 (unpublished) and private
  communication.

\bibitem{stripes}
V.~J. Emery and S.~A. Kivelson, cond-mat/9809083 (to be published in J. of
  Supercond.).

\bibitem{SO5}
E. Demler and S.-C. Zhang, Phys. Rev. Lett. {\bf 75},  4126  (1995).

\bibitem{Zhang2}
E. Demler and S.-C. Zhang, Nature {\bf 396},  733  (1998).

\bibitem{Si}
Q.~M. Si, Y.~Y. Zha, K. Levin, and J.~P. Lu, Phys. Rev. B {\bf 47}, 9055
  (1993). In LaSrCuO, $\epsilon_p-\epsilon^0_d=4$eV, $V_{pd}=0.6$eV, and
  $t_{pp}=-0.15$eV; in YBCO$_{6.6}$, $\epsilon_p-\epsilon^0_d=5$eV,
  $V_{pd}=1.29$eV, $t_{pp}=1.2$eV, and $t'_{pp}=-1.0$eV.

\bibitem{Zha}
Y. Zha, K. Levin, and Q.~M. Si, Phys. Rev. B {\bf 47},  9124  (1993).

\bibitem{Liu}
D.~Z. Liu, Y. Zha, and K. Levin, Phys. Rev. Lett. {\bf 75},  4130  (1995).

\bibitem{earlySi}
Q. Si {\it et~al.}, Physica C {\bf 162-164},  1467  (1989).

\bibitem{Littlewood}
{P. B. Littlewood \textit{et~al.}}, Phys. Rev. B {\bf 48},  487  (1993).

\bibitem{Tremblay}
P. Benard, L. Chen, and M.-M.~S. Tremblay, Phys. Rev. B {\bf 47},  15217
  (1992).

\bibitem{Bulut}
N. Bulut and D.~J. Scalapino, Phys. Rev. B {\bf 53},  5149  (1996).

\bibitem{Mazin}
I.~I. Mazin and V.~M. Yakovenko, Phys. Rev. Lett. {\bf 75},  4134  (1995).

\bibitem{Bulut2}
N. Bulut and D.~J. Scalapino, Phys. Rev. B {\bf 50},  16078  (1994).

\bibitem{Fukuyama}
T. Tanamoto, H. Kohno, and H. Fukuyama, J. Phys. Soc. Japan {\bf 61},  1886
  (1992).

\bibitem{Lee2}
J. Brinckmann and P.~A. Lee, Phys. Rev. Lett. {\bf 82}, 2915 (1999); Related
  calculations were independently reported by Qimiao Si and K. Levin
  (unpublished, 1998).

\bibitem{Yin}
L. Yin, S. Chakravarty, and P.~W. Anderson, Phys. Rev. Lett. {\bf 78},  3559
  (1997).

\bibitem{tetra}
J. Rath {\it et~al.}, Phys. Rev. B {\bf 11},  2109  (1975).

\bibitem{Keimer}
{H. Fong \textit{et~al.}}, cond-mat/9902262 (unpublished).

\bibitem{Lavagna}
M. Lavagna and G. Stemmann, Phys. Rev. B {\bf 49},  4235  (1994).

\bibitem{Keimer2}
{H. F. Fong \textit{et~al.}}, Phys. Rev. Lett. {\bf 75},  316  (1995).

\bibitem{Morr}
{D. Morr and D. Pines }, cond-mat/9807214 (unpublished).

\bibitem{Dai}
P. Dai, H.~A. Mook, and F. Do$\breve{g}$an, Phy. Rev. Lett. {\bf 80},  1738
  (1998).

\bibitem{Yamada}
K. Yamada {\it et~al.}, Phys. Rev. Lett. {\bf 75},  1626  (1995).

\end{thebibliography}

\end{document}